\documentclass[12pt]{article}
\usepackage{amsmath}
\usepackage{graphicx}
\usepackage{subfigure}
\usepackage{hyperref}

\begin{document}

\author{Ernst Trojan \and \textit{Moscow Institute of Physics and Technology} \and 
\textit{PO Box 3, Moscow, 125080, Russia}}
\title{Tachyon stars}
\maketitle

\begin{abstract}
We consider a self-gravitating body composed of ideal Fermi gas of tachyons
at zero temperature. The Oppenheimer-Volkoff equation is solved for various
central densities and various tachyon mass parameter $m$. Although a pure
tachyon star has finite mass, it cannot occur in nature because the
equilibrium condition $P=0$ and the causality condition cannot be satisfied
simultaneously. A stable configuration with tachyon content must be covered
with a non-tachyon envelope. The boundary between the tachyon core and the
envelope is determined by the critical pressure $P_T$, which depends on the
tachyon mass $m$. The tachyon core is dominant and its mass can exceed many
times the solar mass $M_{\odot \,}$ when $m$ is much smaller than the
nucleon mass $m_p$, while at large $m$ compared with $m_p$, the main
contribution to the total stellar mass is due to the envelope whose material
determines the parameters of the whole star. However, the parameters of the
tachyon core do not depend on the envelope material. When the tachyon core
appears, its mass $M_T$ and radius $r_T$ grow up with increasing central
density until maximum values are reached, after which the mass and radius
slowly decrease. The redshift at the surface of the tachyon core does not
depend on $m$ and never exceeds $z_{\max }\simeq 0.3$. The maximum mass of
tachyon core and its maximum radius are achieved at certain central density
and obey universal formulas $M_{T\max }/M_{\odot \,}=0.52m_p^2/m^2$ and $%
r_{T\max }[\mathrm{km}]=4.07\mathrm{\,}m_p^2/m^2$ that allow to estimate
arbitrary supermassive tachyonic bodies at the cosmological scale.
\end{abstract}

\section{Introduction}

Tachyons are instabilities of the field theory with the energy spectrum: 
\begin{equation}
\varepsilon _k=\sqrt{k^2-m^2}  \label{tah}
\end{equation}
where $m$ is the tachyon mass. Tachyons are usually considered in
cosmological models \cite{S2002,BBS2003,FKS02,D1,D2}, and a system of many
tachyons is studied in the frames of statistical mechanics \cite
{M84,DHR89,KRS07}. An ensemble of many tachyons can be considered as a
stable continuous medium when it satisfies the causality 
\begin{equation}
c_s^2=\frac{dP}{dE}\leq 1  \label{ca}
\end{equation}
that implies that the sound speed $c_s$ is subluminal, and that the
functional dependence $P\left[ E\right] $ between the pressure $P$ and
energy density $E$, called as the equation of state (EOS), reveals 'good'
behavior. Otherwise, the system will not be stable with respect to sound
perturbations. It implies that no stable tachyon matter can have free
surface with $P=0$. Thus, if we consider a static self-gravitating
macroscopic tachyon body, it must be embedded in a non-tachyon envelope.
Since the tachyon matter is often discussed in cosmological problems, and it
is important to know whether macroscopic tachyonic objects can exist and
what is their maximum mass. If the mass of the body is great enough, its
gravitational collapse cannot be counterbalanced by the pressure, and it is
necessary to establish the upper bound for the mass until the body becomes a
black hole.

In the present paper we consider a tachyon Fermi gas at zero temperature
which satisfies the causality (\ref{ca}) when its energy density and
pressure exceed critical values \cite{TV2011c} 
\begin{equation}
E_T=\frac{\gamma m^4}{16\pi ^2}\left[ \sqrt{3}-\ln \left( \sqrt{\frac 32}%
+\frac 1{\sqrt{2}}\right) \right]  \label{et}
\end{equation}
\begin{equation}
P_T=\frac{\gamma m^4}{16\pi ^2}\left[ \sqrt{3}+\ln \left( \sqrt{\frac 32}%
+\frac 1{\sqrt{2}}\right) \right]  \label{pt}
\end{equation}
corresponding to the critical Fermi momentum 
\begin{equation}
k_T=\sqrt{\frac 32}m  \label{kt}
\end{equation}
where $\gamma =1$ is the tachyon degeneracy factor. The tachyon pressure
never turns to zero, and tachyon matter cannot have free surface, it must be
dressed in non-tachyon envelope.\textrm{\ }The tachyon matter is endowed
with another peculiar property: its pressure decreases when its energy
density grows up (see Fig.~\ref{tsa0}), and ultra-relativistic EOS 
\begin{equation}
P=\frac E3  \label{ul}
\end{equation}
is achieved at large energy density (when $k_F\gg m$), while the tachyon EOS
becomes 'absolute stiff' 
\begin{equation}
P=E  \label{sti}
\end{equation}
and even 'hyperstiff' 
\begin{equation}
P>E  \label{hyp}
\end{equation}
at low density when $k_F\rightarrow k_T$ (particularly, $P_T=2.23E_T$). So,
a compact cosmic object with tachyon content may be endowed with strange
properties. We consider the simplest example of a static spherical\textrm{\ }%
self-gravitating body with tachyon core and non-tachyon envelope, calculate
its parameters and find the maximum mass of the tachyon core.

Standard relativistic units $c_{light}=\hbar =G=1$ are used in the article.

\section{Tachyon Fermi gas}

The thermodynamical functions of a system of free particles with the
tachyonic energy spectrum $\varepsilon _k$ (\ref{tah}) are determined by
standard formulas of statistical mechanics. Particularly, the energy density
and pressure of a tachyon Fermi gas at zero temperature are \cite{TV2011c}: 
\begin{equation}
E=\frac{\gamma m^4}{8\pi ^2}\left[ \beta ^3\sqrt{\beta ^2-1}-\frac 12\beta 
\sqrt{\beta ^2-1}-\frac 12\ln \left( \beta +\sqrt{\beta ^2-1}\right) \right]
\label{e}
\end{equation}
and 
\begin{equation}
P=\frac{\gamma m^4}{8\pi ^2}\left[ \frac 13\beta ^3\sqrt{\beta ^2-1}+\frac
12\beta \sqrt{\beta ^2-1}+\frac 12\ln \left( \beta +\sqrt{\beta ^2-1}\right)
\right]  \label{p}
\end{equation}
where dimensionless variable 
\begin{equation}
\beta =\frac{k_F}m  \label{b}
\end{equation}
depends on the Fermi-momentum $k_F$ which is linked with the particle number
density according to standard expression \cite{T2011j} 
\begin{equation}
n=\frac{\gamma k_F^3}{6\pi ^2}  \label{n}
\end{equation}
so that the density is given in the form 
\begin{equation}
\rho =mn=\frac{\gamma mk_F^3}{6\pi ^2}=\frac{\gamma m^4}{6\pi ^2}\beta ^3
\label{m}
\end{equation}

It is convenient to express all parameters of tachyon matter in the unit of
normal nuclear density 
\begin{equation}
\rho _0=m_pn_0=2.7\times 10^{14}\,\mathrm{g\cdot cm}^{-3}=157\,\mathrm{%
MeV\cdot fm}^{-3}  \label{nu}
\end{equation}
where $m_p$ is the nucleon mass and 
\begin{equation}
n_0=\frac{4\left( 1.36\,\mathrm{fm}^{-1}\right) ^3}{6\pi ^2}\cong 0.17\,%
\mathrm{fm}^{-3}  \label{nu9}
\end{equation}
is the saturation particle number density of symmetric nuclear matter \cite
{W90}. Thus, dividing (\ref{e}), (\ref{p}) and (\ref{m}) by (\ref{nu}), we
have 
\begin{equation}
\frac E{\rho _0}=\frac{\gamma m^3}{8\pi ^2n_0}\frac m{m_p}\left[ \beta ^3%
\sqrt{\beta ^2-1}-\frac 12\beta \sqrt{\beta ^2-1}-\frac 12\ln \left( \beta +%
\sqrt{\beta ^2-1}\right) \right]  \label{e1}
\end{equation}
\begin{equation}
\frac P{\rho _0}=\frac{\gamma m^3}{8\pi ^2n_0}\frac m{m_p}\left[ \frac
13\beta ^3\sqrt{\beta ^2-1}+\frac 12\beta \sqrt{\beta ^2-1}+\frac 12\ln
\left( \beta +\sqrt{\beta ^2-1}\right) \right]  \label{p1}
\end{equation}
\begin{equation}
\frac \rho {\rho _0}=\frac{mn}{m_pn_0}=\frac{\gamma k_F^3}{6\pi ^2n_0}\frac
m{m_p}=\frac{\gamma m^3}{6\pi ^2n_0}\frac m{m_p}\beta ^3  \label{m1}
\end{equation}

Momentum $k=1\,\mathrm{fm}^{-1}$ corresponds to energy $\varepsilon
_k=k=198\,\mathrm{MeV}$: 
\begin{equation}
1\,\mathrm{fm}^{-1}\leftrightarrow 198\,\mathrm{MeV}  \label{nuu}
\end{equation}
while $k=4.74\,\mathrm{fm}^{-1}$ corresponds to the mass of nucleon\textrm{\ 
}$m_p=939\,\mathrm{MeV}$, and normal nuclear density (\ref{nu}) corresponds
to 
\begin{equation}
\rho _0\leftrightarrow 0.80\,\mathrm{fm}^{-4}  \label{nu2}
\end{equation}
The tachyon mass parameter $m$ plays the role of form-factor in equations (%
\ref{e1})-(\ref{m1}). Taking $m=4.74\bar m=4.74m/m_p$, we find that 
\begin{equation}
\frac{m^3}{2\pi ^2n_0}\frac m{m_p}=\frac{\left( 4.74\right) ^3}{2\pi ^2\cdot
0.17}\frac{m^4}{m_p^4}=31.74\bar m^4  \label{dim}
\end{equation}
Hence, substituting (\ref{dim}) in (\ref{e1})-(\ref{m1}), we express the
energy density and pressure in nuclear units:

\begin{equation}
\frac E{\rho _0}=7.93\gamma \bar m^4\left[ \beta ^3\sqrt{\beta ^2-1}-\frac
12\beta \sqrt{\beta ^2-1}-\frac 12\ln \left( \beta +\sqrt{\beta ^2-1}\right)
\right] \equiv \bar E\bar m^4  \label{e2}
\end{equation}
\begin{equation}
\frac P{\rho _0}=7.93\gamma \bar m^4\left[ \frac 13\beta ^3\sqrt{\beta ^2-1}%
+\frac 12\beta \sqrt{\beta ^2-1}+\frac 12\ln \left( \beta +\sqrt{\beta ^2-1}%
\right) \right] \equiv \bar P\bar m^4  \label{p2}
\end{equation}
\begin{equation}
\frac \rho {\rho _0}=10.58\gamma \bar m^4\beta ^3\equiv \bar \rho \bar m^4
\label{m2}
\end{equation}
where dimensionless $\bar E$, $\bar P$ and $\bar \rho $ imply the energy
density, pressure and density of tachyon gas with $m=m_p$.

By the way, the well-known EOS of the ordinary Fermi gas of subluminal
particles with the energy spectrum 
\begin{equation}
\varepsilon _k=\sqrt{k^2+m^2}  \label{bra}
\end{equation}
can be given in the universal form 
\begin{equation}
E=\frac{\gamma m^4}{8\pi ^2}\left[ \beta ^3\sqrt{\beta ^2+1}-\frac 12\beta 
\sqrt{\beta ^2+1}-\frac 12\ln \left( \beta +\sqrt{\beta ^2+1}\right) \right] 
\label{e3}
\end{equation}
\begin{equation}
P=\frac{\gamma m^4}{8\pi ^2}\left[ \frac 13\beta ^3\sqrt{\beta ^2+1}+\frac
12\beta \sqrt{\beta ^2+1}+\frac 12\ln \left( \beta +\sqrt{\beta ^2+1}\right)
\right]   \label{p3}
\end{equation}
that also includes dimensionless variable $\beta $ (\ref{b}) and the same
form-factor $\bar m^4$. 
Both tachyon and bradyon EOS tend to the same
ultra-relativistic limit 
\begin{equation}
P=\frac E3\simeq \frac{\gamma m^4}{24\pi ^2}\beta ^4\sim n^{4/3}  \label{po1}
\end{equation}
when $\beta \rightarrow \infty $. However the tachyon EOS reveals the
following non-relativistic ($\beta \rightarrow 1$) behavior 
\begin{equation}
P\rightarrow \frac{\gamma \sqrt{2}m^4}{4\pi ^2}\sqrt{\beta -1}=\frac{\gamma
m^4}{4\pi ^2}\sqrt{\frac 23\left( \frac n{n_0}-1\right) }\gg E\rightarrow 
\frac{\gamma \sqrt{2}m^4}{3\pi ^2}\left( \sqrt{\beta -1}\right) ^3=\frac{%
\gamma m^4}{3\pi ^2}\sqrt{\frac 23\left( \frac n{n_0}-1\right) ^3}\qquad
n\rightarrow \frac{\gamma m^3}{6\pi ^2}  \label{po2}
\end{equation}
while the EOS of non-relativistic ($\beta \rightarrow 0$) bradyons obeys
standard formulas 
\begin{equation}
P\rightarrow \frac{\gamma m^4\beta ^5}{30\pi ^2}=\frac{\left( 6\pi ^2/\gamma
\right) ^{2/3}}{5m}n^{5/3}\qquad E\rightarrow mn+\frac 32P\qquad
n\rightarrow 0\sqrt{\frac 23\left( \frac{6\pi ^2n}\gamma \left( \frac
{}{}\right) -1\right) }\qquad n=1+\frac 13\frac x{n_0}=\beta   \label{po3}
\end{equation}

Dependence of ratio $P/E$ vs $\beta $ is given in Fig.~\ref{tsa0}. The EOS
of cold tachyon Fermi (\ref{e2})-(\ref{p2}) is hyperstiff (\ref{hyp}) when 
\cite{TV2011c} 
\begin{equation}
\beta <\beta _1=1.529  \label{k1}
\end{equation}
However, the variable $\beta $ cannot be arbitrary small because the
tachyonic parameters (\ref{e})-(\ref{p}) satisfy the causality condition (%
\ref{ca}) when the Fermi momentum exceeds the critical value (\ref{kt})
corresponding to 
\begin{equation}
\beta \geq \beta _T=\sqrt{\frac 32}\cong 1.225  \label{bt}
\end{equation}
\textrm{\ }and no stable tachyon matter exists at low density 
\begin{equation}
\rho <\rho _T=\frac{\gamma m^4}{6\pi ^2}\left( \frac 32\right) ^{3/2}
\label{mt}
\end{equation}
while the energy density and pressure cannot fall below critical minimum
values (\ref{et}) and (\ref{pt}). The critical parameters (\ref{mt}), (\ref
{et}) and (\ref{pt}) are 
\begin{equation}
\frac{E_T}{\rho _0}=4.26\gamma \bar m^4  \label{e2t}
\end{equation}
\begin{equation}
\frac{P_T}{\rho _0}=9.48\gamma \bar m^4  \label{p2t}
\end{equation}
\begin{equation}
\frac{\rho _T}{\rho _0}=19.44\gamma \bar m^4  \label{m2t}
\end{equation}

The critical density (\ref{m2t}) differs from the value $\rho _T$ obtained
in the previous research \cite{TV2011c} where the tachyon particle number
density was defined as 
\begin{equation}
n=\frac \gamma {6\pi ^2}\left( k_F^3-m^3\right)  \label{mo}
\end{equation}
that is not correct \cite{T2011j}. Now we use the correct definition (\ref{n}%
) that yields zero entropy of the cold tachyon Fermi gas in agreement with
the Nernst heat theorem because $E+P-\varepsilon _Fn=0$, according to (\ref
{e}), (\ref{p}) and (\ref{m}), where $\varepsilon _F=\sqrt{k_F^2-m^2}$ is
the Fermi energy of tachyon gas.

\section{Properties of pure tachyon star}

A stable spherical-symmetric configuration of self-gravitating body is
determined by the Oppenheimer-Volkoff equation \cite{OV39}: 
\begin{equation}
\frac{dP}{dr}=-(E+P)\frac{M+4\pi r^3P}{r\left( r-2M\right) }  \label{o1}
\end{equation}
where 
\begin{equation}
\frac{dM}{dr}=4\pi r^2E  \label{o2}
\end{equation}
and $M\left( r\right) $ is the mass distribution along radius $r$. Equations
(\ref{o1})-(\ref{o2}) must be solved under the initial conditions $M\left(
0\right) =0$ and $E_{*}=E\left( 0\right) $. The latter can be formulated in
the form $P_{*}=P\left( 0\right) $ or $\rho _{*}=\rho \left( 0\right) $. The
boundary condition 
\begin{equation}
P\left( r_{*}\right) =0  \label{bdr}
\end{equation}
determines the total mass of the body $M_{*}=M\left( r_{*}\right) $ and its
radius $r_{*}$. Therefore, parameters, $M_{*}$ and $r_{*}$ depend on the EOS
and the central density $\rho _{*}$.

Equations (\ref{o1})-(\ref{o2}) allow to determine the upper bound of the
mass until the body becomes a black hole.\textrm{\ }In general, the stiffer
the EOS, the greater the mass of the star. The Oppenheimer-Volkoff limit $%
M_{*}=0.71M_{\odot }$ is obtained with non-interacting neutron gas. This
limit becomes much greater for interacting neutron-rich matter,
particularly, $M_{*}=1.64M_{\odot }$ for the EOS calculated in the frames of
nonlinear sigma-model \cite{TV2010}. The maximum possible stellar mass $%
M_{*}=2.9\div 3.2M_{\odot }$ \cite{RR74,KB96} is obtained when the 'absolute
stiff' EOS (\ref{sti}) is applied\textrm{. }

However, a self-gravitating body composed of the tachyon matter may reveal
unexpected 'hyperstiff' behavior (\ref{hyp}), see Fig.~\ref{tsa0}. The tachyon
matter in the dense center of the star will be 'soft', and its 'stiffness'
will increase from the center to the edges where the density is smaller.
Contrary to the gas of subluminal massive particles, the tachyon matter at
low density (when $\beta =k_F/m\rightarrow 1$) cannot be described by simple
polytrope (\ref{po1})-(\ref{po2}). It implies that a tachyon compact stellar
object may have much more greater mass than the stars composed of regular
matter and even 'absolute stiff' matter.

Substituting (\ref{e2}) and (\ref{p2}) in (\ref{o1})-(\ref{o2}), we write equations in the dimensionless form 
\begin{equation}
\frac{d\bar P}{dr}=-0.148(\bar E+\bar P)\frac{M+1.77\bar m^4r^3\bar P}{%
r\left( r-0.295M\right) }  \label{o1a}
\end{equation}
\begin{equation}
\frac{dM}{dr}=1.77\bar m^4r^2\bar E  \label{o2a}
\end{equation}
where $M$ is expressed in the unit of solar mass $M_{\odot }=1.99\times
10^{33}~\mathrm{g}$, and $r$ is expressed in $10~\mathrm{km}$. Here the
dimensionless form-factor $\bar m=m/m_p$ plays the role of scaling.
Substituting 
\begin{equation}
\bar M=M\bar m^2\qquad \bar r=r\bar m^2  \label{sc0}
\end{equation}
in (\ref{o1a})-(\ref{o2a}) we rewrite the equations so 
\begin{equation}
\frac{d\bar P}{d\bar r}=-0.148(\bar E+\bar P)\frac{\bar M+1.77\bar r^3\bar P%
}{\bar r\left( \bar r-0.295\bar M\right) }  \label{o1b}
\end{equation}
\begin{equation}
\frac{d\bar M}{d\bar r}=1.77\bar r^2\bar E  \label{o2b}
\end{equation}
where $\bar M$ is expressed in the unit of $M_{\odot }\bar m^2$, and $\bar r$
is expressed in $10\bar m^2~\mathrm{km}$.

So, we need to simulate equations (\ref{o1b})-(\ref{o2b}) in order to find
the total mass $\bar M_{*}$ and radius $\bar r_{*}$, while the stellar
parameters at arbitrary $m=\bar mm_p$ will be automatically determined,
according to scaling (\ref{sc0}), so 
\begin{equation}
M_{*}=\frac{\bar M_{*}}{\bar m^2}\qquad r_{*}=\frac{\bar r_{*}}{\bar m^2}
\label{sc}
\end{equation}

The theory states that the maximal mass of neutron stars is achieved when
the central density $\rho _{*}=5\div 10\rho _0$ \cite{TV2010,KB96},
depending on the particular model of nuclear matter. The relevant central
density in a tachyon self-gravitating body may vary in much more wider
range, depending on the choice of the tachyon mass $m$. So, it is convenient
to operate with dimensionless Fermi momentum $\beta $ (\ref{b}) rather than
tachyon density $\rho =10.58\rho _0\bar m^4\beta ^3$ (\ref{m2}). Now we need
to find only one profile $\bar M_{*}$ vs $\beta _{*}$ in order to build all
other profiles at different $m$, applying rescaling of the mass $M_{*}$ (\ref
{sc}).

The similar scaling (\ref{sc0}) is applied to the Oppenheimer-Volkoff
equation when operate with the EOS of cold Fermi gas of subluminal particles 
\cite{OV39}. Of course, we can immediately calculate an ordinary neutron
star with pure neutron content, substituting the EOS of subluminal Fermi gas
(\ref{e3})-(\ref{p3}) in (\ref{o1b})-(\ref{o2b}). This star will always have
finite mass and finite radius because the pressure turns to zero at finite%
\textrm{\ }$\bar r_{*}$ (which depends on the central parameter $\beta _{*}$
and form-factor $m$). Indeed, it is due to the behavior of tachyonic
thermodynamical functions (\ref{po2})-(\ref{po3}) at $\beta \rightarrow 1$
(see Fig.~\ref{tsa0}).

Results of calculation for a pure tachyon body are given in Fig.~\ref{tsa1}-%
\ref{tsa5}. The stellar mass $\bar M_{*}$ is finite (Fig.~\ref{tsa1}), while
its radius tends to infinity because the hydrostatic equilibrium $\bar
P\left( \bar r_{*}\right) =0$ is achieved when $\bar r_{*}\rightarrow \infty 
$. In fact the tachyonic pressure never turns to zero.

So, we can estimate an effective radius $\bar r_{90}$ together with $\bar
r_{99}$ under which only $90\%$ and $99\%$ of the total mass is enclosed,
respectively (dashed and dotted vertical lines in Fig.~\ref{tsa1}). The
effective radius of tachyon star remains almost constant 
\begin{equation}
\bar r_{*}\equiv \bar r_{99}\simeq 50~\mathrm{km}  \label{r0}
\end{equation}
when the dimensionless Fermi momentum is large $\beta _{*}>2$ (Fig.~\ref
{tsa4}), and $\bar r_{99}$ increases, tending to infinity, when $\beta
_{*}\rightarrow 1$ .

The maximum mass of tachyon star 
\begin{equation}
\bar M_{*\max }=1.18M_{\odot }  \label{m0}
\end{equation}
is achieved at 
\begin{equation}
\beta _{*}=1.81  \label{b0}
\end{equation}
corresponding to the central density 
\begin{equation}
\rho _{*}=62.7\rho _0  \label{rho0}
\end{equation}
see Fig.~\ref{tsa2} and Fig.~\ref{tsa3}).

Relation (\ref{sc0}) implies that the redshift 
\begin{equation}
z=\frac 1{\sqrt{1-2M/r}}-1=\frac 1{\sqrt{1-0.295\bar M/\bar r}}-1
\label{red}
\end{equation}
does not depend on the tachyon mass $m$. The redshift attains its maximum $%
z_{\max }$ inside the star and decreases closer to the surface (Fig.~\ref
{tsa5}). The value $z_{\max }$ increases with increase of the central
parameter $\beta _{*}$, tending to $z_{\max }\simeq 0.41$ at $\beta
_{*}\rightarrow \infty $. Substituting the effective radius (\ref{r0}) in (%
\ref{red}), we can define the effective surface redshift 
\begin{equation}
z_{99}=0.036  \label{red2}
\end{equation}
associated with $\bar r_{99}$ (\ref{r0}) that remains the same without
regard of $m$.

We have determined the parameters of tachyon star at $\bar m=m/m_p=1$, and,
according to rescaling (\ref{sc}) we can immediately calculate the stellar
parameters at arbitrary $m$, namely the maximum mass
\begin{equation}
M_{*\max }=\frac{1.18M_{\odot }}{\bar m^2}  \label{m00}
\end{equation}
is achieved at the central density 
\begin{equation}
\rho _{*\max }=62.7\rho _0\bar m^4  \label{d00}
\end{equation}
corresponding to the same $\beta _{*\max }=1.81$ (\ref{b0}), while the
effective radius of the star is 
\begin{equation}
r_{*}\equiv r_{99}=\frac{50~\mathrm{km}}{\bar m^2}  \label{r00}
\end{equation}
For example, at $m=266~\mathrm{MeV}$ the maximum mass of tachyon star is 
\begin{equation}
M_{*}=14.7M_{\odot }  \label{m02}
\end{equation}
and its effective radius is 
\begin{equation}
r_{*}\equiv r_{99}=623~\mathrm{km}  \label{r02}
\end{equation}
The maximal mass (\ref{m02}) is achieved at the central density $\rho
_{*}=0.41\rho _0$.

According to equations (\ref{o1b})-(\ref{o2b}), we have calculate
self-gravitating body composed of pure tachyon matter, whose density is
considered to be arbitrary low at the surface, when $P\rightarrow 0$ at $%
r\rightarrow \infty $. Meanwhile the tachyon matter does not satisfy the
causality (\ref{ca}) when $\beta _F<\beta _T=\sqrt{3/2}$ (\ref{kt}) and $%
P<P_T$ (\ref{pt}). It implies that no stable stellar configuration of pure
tachyon matter is possible in practice. So, we need to revise the results in the light of realistic model which could have practical significance.

\section{Tachyon core in non-tachyonic envelope}

The stable tachyon matter cannot have free surface with $P=0$ because it
cannot satisfy the causality (\ref{ca}) at arbitrary small density $\rho $,
smaller than $\rho _T$ (\ref{mt}), and its pressure cannot fall below the
critical value $P_T$ (\ref{pt}). Therefore, if the tachyon matter appears in
some macroscopic domain, it must be dressed in a non-tachyonic envelope
where hydrostatic equilibrium (\ref{bdr}) could be achieved. Hence, the
radius of such stellar configuration $r_{*}$\ will be greater than the
radius of tachyon core $r_T$ determined from condition $P\left( r_T\right)
=P_T$. The mass of tachyon core $M_T=M\left( r_T\right) $ will be smaller
than the total mass of the body $M_{*}=M\left( r_{*}\right) $ where $r_{*}$
is found from condition $P\left( r_{*}\right) =0$ (\ref{bdr}).

Let us analyze a stable configuration of tachyonic self-gravitating body
immersed in some non-tachyonic medium. The tachyon core exists in the
central region where the pressures exceeds the critical value $P>P_T$ (\ref
{p2t}), corresponding to density $\rho >\rho _T$ (\ref{m2t}) and Fermi
momentum $k>k_T$ (\ref{kt}). The non-tachyonic envelope exists at $P<P_T$,
and the boundary between the tachyon core and the envelope is determined as 
\begin{equation}
P_T=P_{env}  \label{edg}
\end{equation}
where the pressure $P_{env}$ is calculated with the particular EOS of the
envelope. (It should be noted that the upper density in the envelope $\rho
_{env}$, in general, does not coincide with $\rho _T$).

We consider three variants of non-tachyonic material. The simplest idea is
to put the tachyon core into a pure neutron (PN) envelope whose EOS is given
by formulas (\ref{e2})-(\ref{p2}) at $m=m_p$. For the envelope composed of
'absolute stiff' (AS) neutron matter its EOS $P=E$ (\ref{sti}) is applied
when $\rho >\rho _s=4.6\times 10^{14}\,\mathrm{g\cdot cm}^{-3}$ and the
Baym-Pethick-Sutherland (BPS) EOS \cite{BPS75} is applied when $\rho <\rho
_s $ \cite{RR74}. Both PN and AS EOS are no more than limiting theoretical
possibilities, while the real nuclear matter includes strong interaction and
its EOS is stiffer than PN EOS but softer than AS EOS. For a reliable
example we consider the neutron-rich matter in the nonlinear sigma model
(NSM) \cite{TV2010,V98}. This NSM EOS was calculated at a high accuracy,
including the effect of correlation energy, and its EOS stands between PN
and AS -- by these three alternatives we manage to embrace the whole range
of possible envelope materials. The BPS EOS is applied when the density in
the envelope becomes smaller than $\rho _{\perp }=2.5\times 10^{14}\,\mathrm{%
g\cdot cm}^{-3}$ (although the outer layers at so small density do not
affect the result).

The mass and radius of this stellar configuration is always finite, and it
may depend on the sort of the envelope. At $\beta _{*}<$ $\sqrt{3/2}$ the
star has no tachyon content and we do not discuss this range of densities in
detail because the parameters of ordinary non-tachyonic neutron stars are
already calculated \cite{TV2010,OV39,RR74,KB96}. We calculate the total mass 
$M_{*}$ and radius $r_{*}$ as well as the mass of tachyon core $M_T$ and its
radius $r_T$ at different dimensionless Fermi momentum $\beta _{*}\geq $ $%
\sqrt{3/2}$ (corresponding to the central density $\rho _{*}$ $\geq \rho _T$%
) and for several values of $m$, see Fig.~\ref{tsa939a}-\ref{tsa88}.

When the tachyon mass equals to the nucleon mass $m=m_p=939\,\mathrm{MeV}$
and the tachyon critical density is $\rho _T=19.44\rho _0$, the total mass
of the star is around $M_{*}\simeq 1.3M_{\odot }$, $M_{*}\simeq 1.4M_{\odot
} $ and $M_{*}\simeq 2.9M_{\odot }$ for the star with PN, NSM and AS
envelope, respectively, while its total radius is around $6\,\mathrm{km}$, $%
9\,\mathrm{km}$, and $16\,\mathrm{km}$ and , see Fig.~\ref{tsa939a} and Fig.~%
\ref{tsa939b}. While the total stellar mass is definitely greater than the
relevant mass of non-tachyon star, the total radius is subject to no visible
growth (the negligible increment round 0.5 km is not depicted in Fig.~\ref{tsa939b}). The envelope makes major contribution to the total stellar mass, and
the sort of material is very important here. The PN envelope is most
sensible and its mass has increased almost three times with respect to the
mass of PN star without the tachyon content.

As for the parameters of the tachyon core, they do not depend on the
envelope material. The tachyon core appears in the star as soon as $\beta
_{*}$ exceeds $\sqrt{3/2}$ and its mass and radius increase with increasing $%
\beta _{*}$ until maximum values $M_T=0.52M_{\odot }$ and $r_T=4.07\,\mathrm{km}$
are reached, after which the mass and radius decrease, see Fig.~\ref{tsa939a}%
. The radius of tachyon core remains finite and it is almost a constant
around $3.5\,\mathrm{km}$ at large $\beta _{*}$ (Fig.~\ref{tsa939b}).


Since the parameters of the tachyon core do not depend on the envelope EOS,
we may choose the simplest variant of PN envelope to investigate the system
at different tachyon mass $m$. At $m=666\,\mathrm{MeV}$ ($\rho _T=4.91\rho
_0 $) the maximum mass of tachyon core is $M_T=1.02M_{\odot }$ and its
maximum radius is $r_T=8.1\,\mathrm{km}$, while the mass of the whole
stellar mass is around $M_{*}=2.4M_{\odot }$, and its radius does not exceed 
$10$~km (see Fig.~\ref{tsa666a}). At $m=400\,\mathrm{MeV}$ ($\rho
_T=0.64\rho _0$) the maximum mass of tachyon core is $M_T=2.85M_{\odot }$
and its maximum radius approaches $22.4\,\mathrm{km}$, while the total
stellar mass is around $M_{*}=6.0M_{\odot }$, and its radius does not exceed 
$25$~km, (see Fig.~\ref{tsa400a}). At $m=233\,\mathrm{MeV}$ ( $\rho
_T=0.07\rho _0$) the tachyon core dominates over the envelope, without
regard of its material (see Fig.~\ref{tsa233a}). The maximum mass of tachyon
core is $M_T=8.28M_{\odot }$ and its maximum radius is expanded up to $%
r_T=66\,\mathrm{km}$. However, the total stellar mass is around $%
M_{*}=9.0M_{\odot }$, and its radius never exceeds $r_{*}=$ $70\,\mathrm{km}$%
. Now the tachyon core plays much more important role and its appearance is
reflected in visible changes of the total stellar parameters, while the
contribution of the envelope is small. At $m=m_p=138\,\mathrm{MeV}$ the mass
of tachyon core becomes as large as $M_T=23.7M_{\odot }$ and its maximum
radius is around $188\,\mathrm{km}$ (see Fig.~\ref{tsa138a}), while the
envelope is thin and light and its contribution is negligible so that $%
M_{*}\simeq M_T$ and $r_{*}\simeq r_T$.

However, the only calculation \ref{tsa138b} is enough because, without
regard of the envelope EOS, all profiles of the tachyon core obey the same
scaling 
\begin{equation}
M_T\left[ \beta _{*}\right] =\frac{\bar M_T\left[ \beta _{*}\right] }{\bar
m^2}\qquad r_T\left[ \beta _{*}\right] =\frac{\bar r_T\left[ \beta
_{*}\right] }{\bar m^2}  \label{sc33}
\end{equation}
where profiles $\bar M_T\left[ \beta _{*}\right] $ and $\bar r_T\left[ \beta
_{*}\right] $\ (see Fig.~\ref{tsa138b}) are calculated at $\bar m$\ $%
=m/m_p=1 $. It should be also noted that the maximum mass $M_{T\max }$\ is
achieved at the same $\beta _{\max }=2.18$, while the maximum radius $%
r_{T\max }$ is achieved at the same $\beta _{*}=1.61$.

The scaling (\ref{sc33}) implies that the redshift at the surface of the
tachyon core is the same for every $m$ (see Fig.~\ref{tsa88}) while the
maximum value $z_{T\max }=0.30$ is achieved at the same $\beta _{*}=2.59$
without regard of $m$.

\section{Conclusion}

\begin{figure}[tbp]
\caption{Ratio pressure/energy vs Fermi momentum $\beta =k_F/m$ for cold
Fermi gases of tachyons (solid) and bradyons (dashed). }
\label{tsa0}{\includegraphics[scale=0.7]{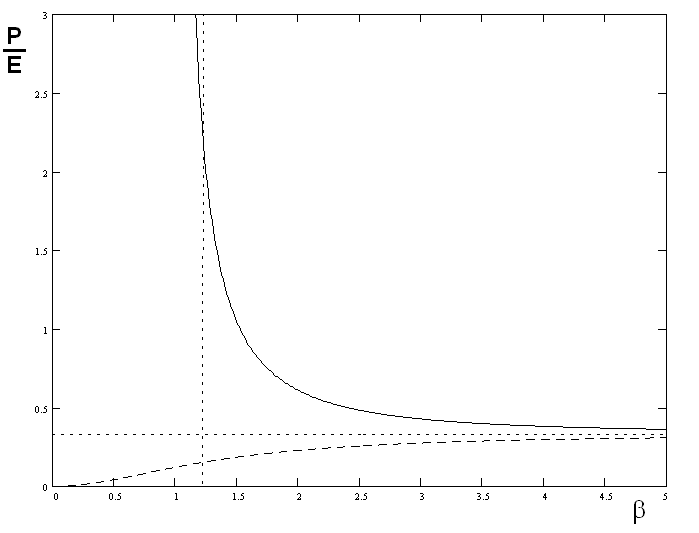}}
\par
The tachyon gas is unstable in the left of vertical dotted line (corresponds
to $\beta<\sqrt{3/2}$). Horizontal dotted line corresponds to
ultrarelativistic gas with $P/E=1/3$.
\end{figure}

The thermodynamical functions of the cold tachyon Fermi gas (\ref{e2})-(\ref
{p2}) include the same form-factor $\sim m^4$ as the EOS of cold Fermi gas
of subluminal particles (\ref{e3})-(\ref{p3}), but the tachyonic EOS may
occur 'hyperstiff' $P>E$ (see Fig.~\ref{tsa0}), and it makes sufficient
difference from the usual analysis of neutron stars.

\begin{figure}[tbp]
\caption{Mass distribution $M(r)$ inside a self-gravitating body of pure
tachyon matter (solid graph) and pure neutron matter at the same $m=939\, 
\mathrm{MeV}$ (dashed). }
\label{tsa1}{\includegraphics[scale=0.8]{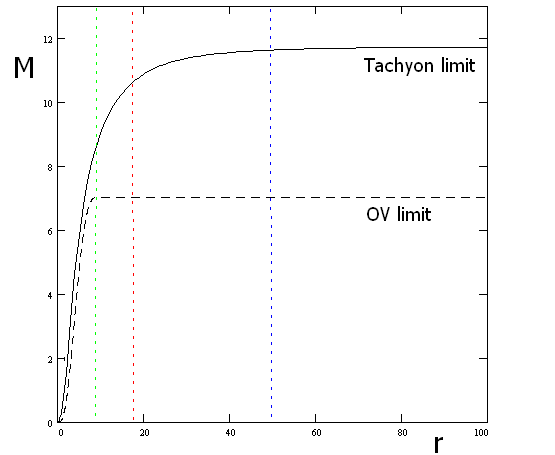}}
\par
The neutron star has finite mass (Oppenheimer-Volkoff limit) and finite
radius (green vertical line), the tachyon star has finite mass and infinite
radius. Blue vertical line cuts 90\%, red vertical line cuts 99\% of the
total mass of tachyon star.
\end{figure}

\begin{figure}[tbp]
\caption{Mass of tachyonic self-gravitating body vs central density. }
\label{tsa2}{\includegraphics[scale=0.7]{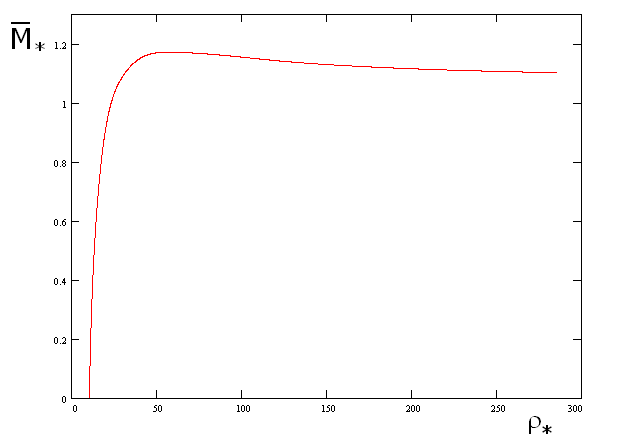}}
\end{figure}

\begin{figure}[tbp]
\caption{Mass of tachyonic self-gravitating body vs variable. }
\label{tsa3}{\includegraphics[scale=0.8]{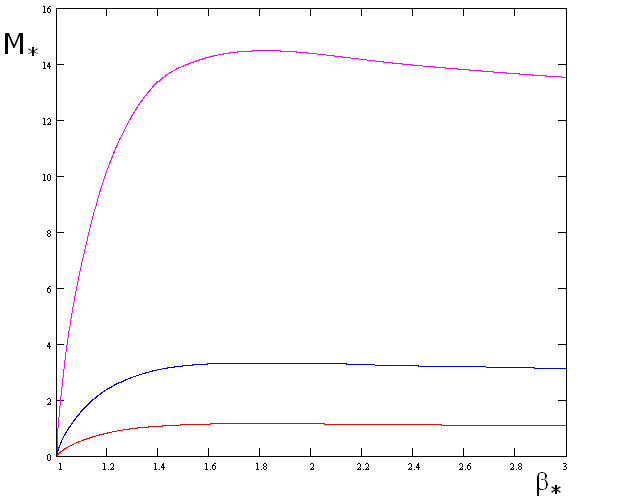}}
\par
At $m=266\, \mathrm{MeV}$ (pink), $m=666\, \mathrm{MeV}$ (blue) and $m=939\, 
\mathrm{MeV}$ (red).
\end{figure}

\begin{figure}[tbp]
\caption{Effective radius $\bar r_{99}$ (red) and $\bar r_{90}$ (blue) of
tachyonic self-gravitating body vs central variable $\beta_*$. }
\label{tsa4}{\includegraphics[scale=0.7]{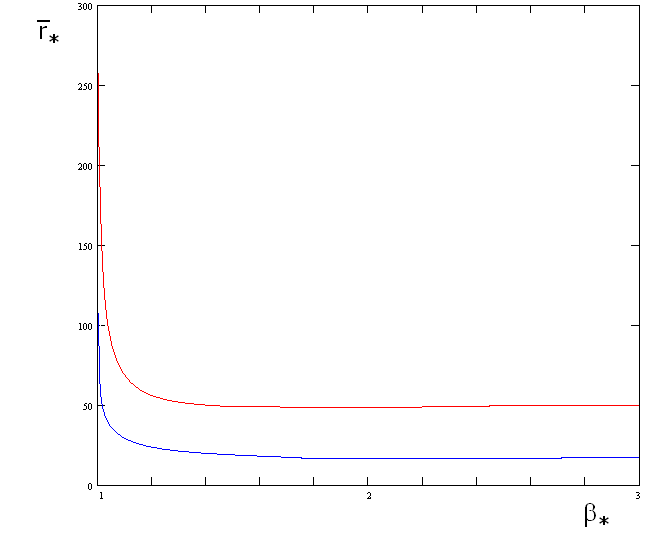}}
\end{figure}

\begin{figure}[tbp]
\caption{The profile of redshift $z(r)$ inside the tachyonic body at
different central density: $\beta_* =1.5$ (solid line), $\beta_* =2.5$
(dashed), $\beta_* =5$ (dotted) and maximum redshift $z_{\mathrm{max}}$ vs $%
\beta_*$. }
\label{tsa5}{\includegraphics[scale=0.5]{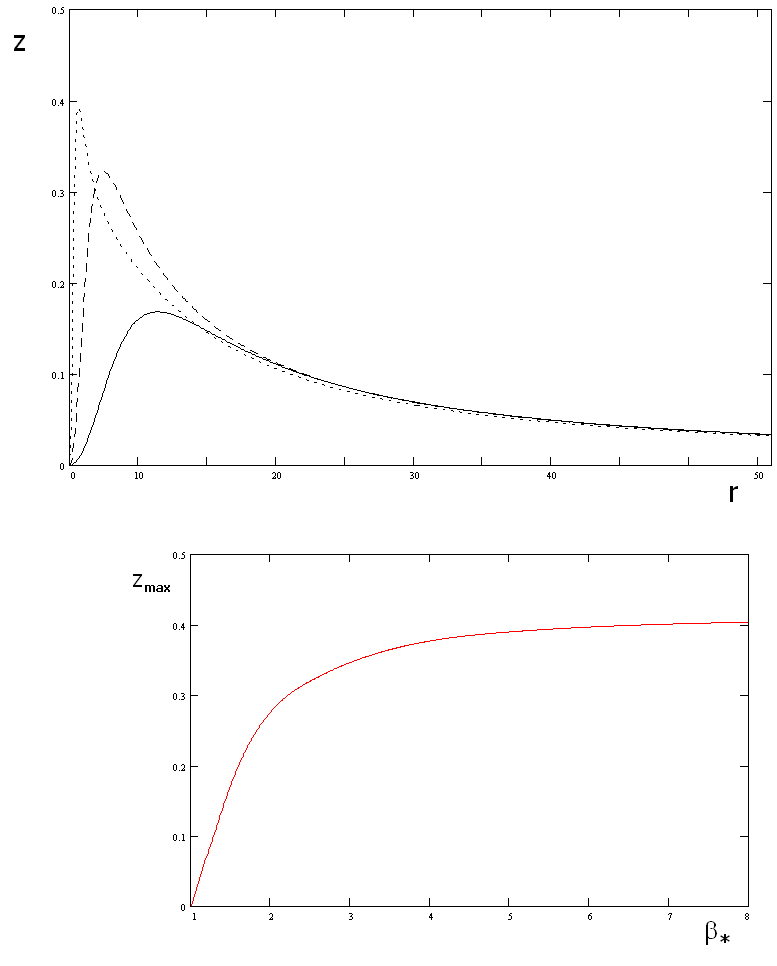}}
\end{figure}

The parameters of tachyonic self-gravitating body depend on the central
density $\rho _{*}$ and the tachyon mass $m$. For a pure tachyon star its
mass and effective radius are plotted in Fig.~\ref{tsa1}-\ref{tsa5},
particularly, the maximum mass $M=1.18M_{\odot }$ is achieved at $m=939~%
\mathrm{MeV}$. The maximum mass of a tachyonic body at arbitrary $m$ is
determined by universal scaling formula (\ref{m00}) and it is achieved at
the same Fermi momentum $k_F=1.81m$ (\ref{m00}). While the size of a star
with pure neutron content is finite \cite{OV39}, the size of pure tachyon
star is unbound. However, we can determine its effective radius (\ref{r00})
under which $99\%$ of the total mass is enclosed. The relevant effective
redshift (\ref{red2}) is constant and does not depend on $m$.

In practice a stable self-gravitating tachyon body can exist only if it is
covered with a non-tachyon envelope because the tachyon matter is unstable
when its pressure is below the critical value $P_T>0$ (\ref{pt}) and
hydrostatic equilibrium condition $P=0$ must be achieved somewhere beyond
the tachyon matter. We have calculated tachyon stars with three sorts of
envelopes: pure neutron matter, neutron-rich matter in the non-linear sigma
model \cite{TV2010,V98} and absolute stiff matter (\ref{sti}).

\begin{figure}[tbp]
\caption{The mass of a star with tachyon core and three types of envelope vs
variable $\beta_*$ (at $m=939\, \mathrm{MeV}$). }
\label{tsa939a}{\includegraphics[scale=0.75]{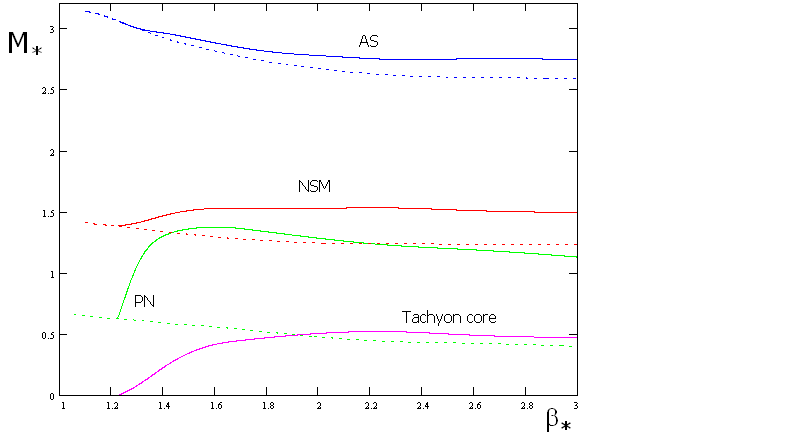}}
\par
Dashed line corresponds to a star with no tachyon content.
\end{figure}

\begin{figure}[tbp]
\caption{The radius of a star with tachyon core and three types of envelope
vs $\beta_*$ ($m=939\, \mathrm{MeV}$). }
\label{tsa939b}{\includegraphics[scale=0.8]{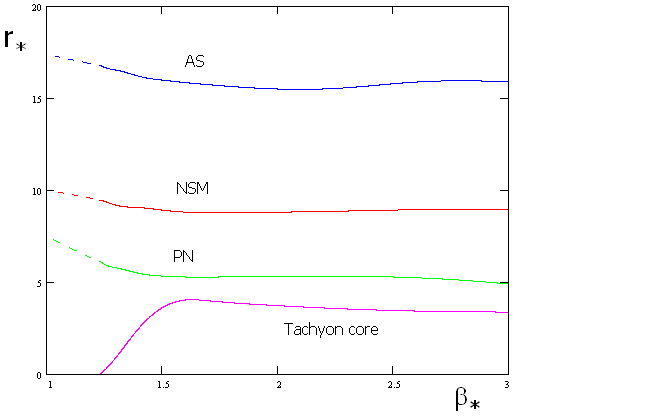}}
\par
Notation is the same as in Fig.~\ref{tsa939a}. The radius of tachyon core is
given in PINK.
\end{figure}

\begin{figure}[tbp]
\caption{The mass and radius of a star with tachyon core in PN envelope vs
variable $\beta_*$ (at $m=666\, \mathrm{MeV}$). }
\label{tsa666a}{\includegraphics[scale=0.52]{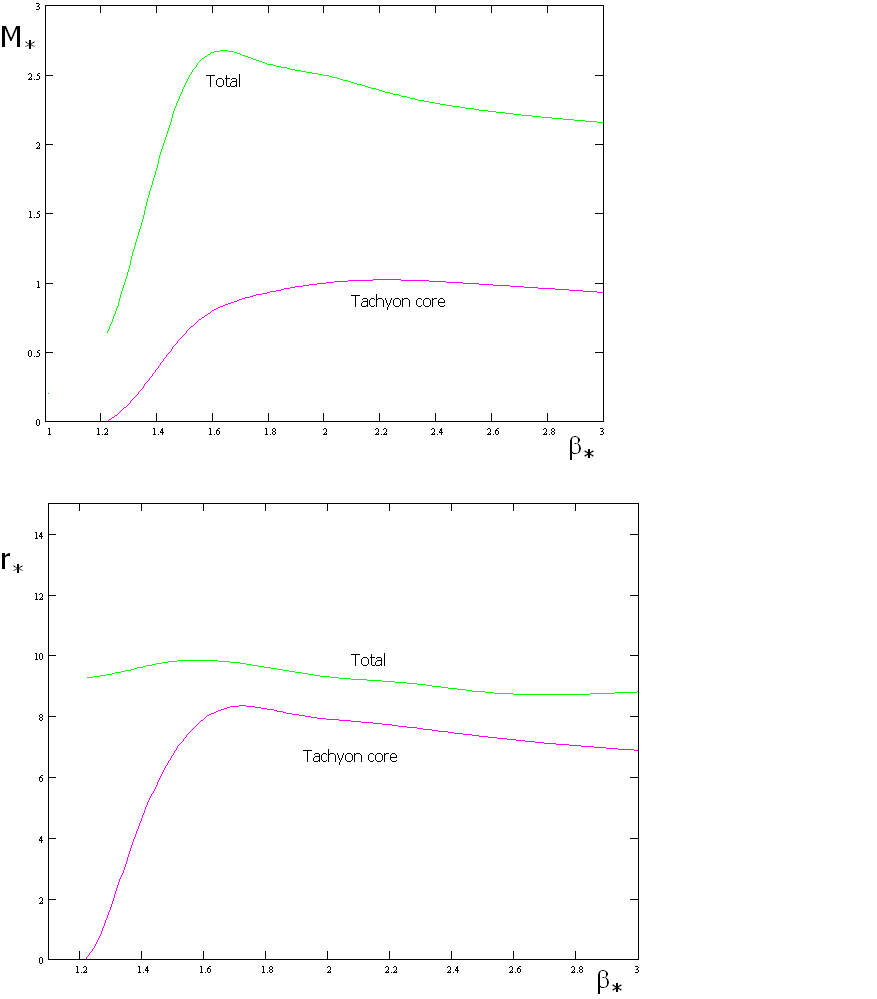}}
\par
\end{figure}

\begin{figure}[tbp]
\caption{The mass and radius of a star with tachyon core in PN envelope vs
variable $\beta_*$ (at $m=400\, \mathrm{MeV}$). }
\label{tsa400a}{\includegraphics[scale=0.55]{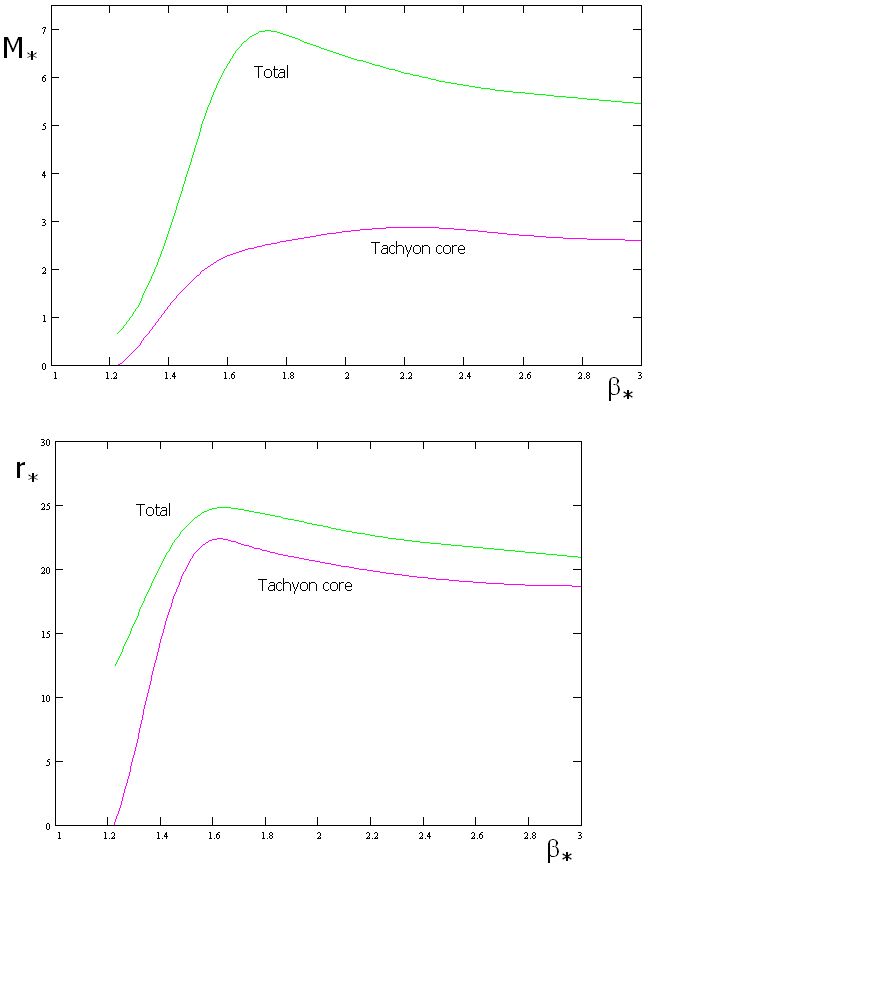}}
\par
\end{figure}

\begin{figure}[tbp]
\caption{The mass and radius of a star with tachyon core in PN envelope vs
variable $\beta_*$ (at $m=233\, \mathrm{MeV}$). }
\label{tsa233a}{\includegraphics[scale=0.5]{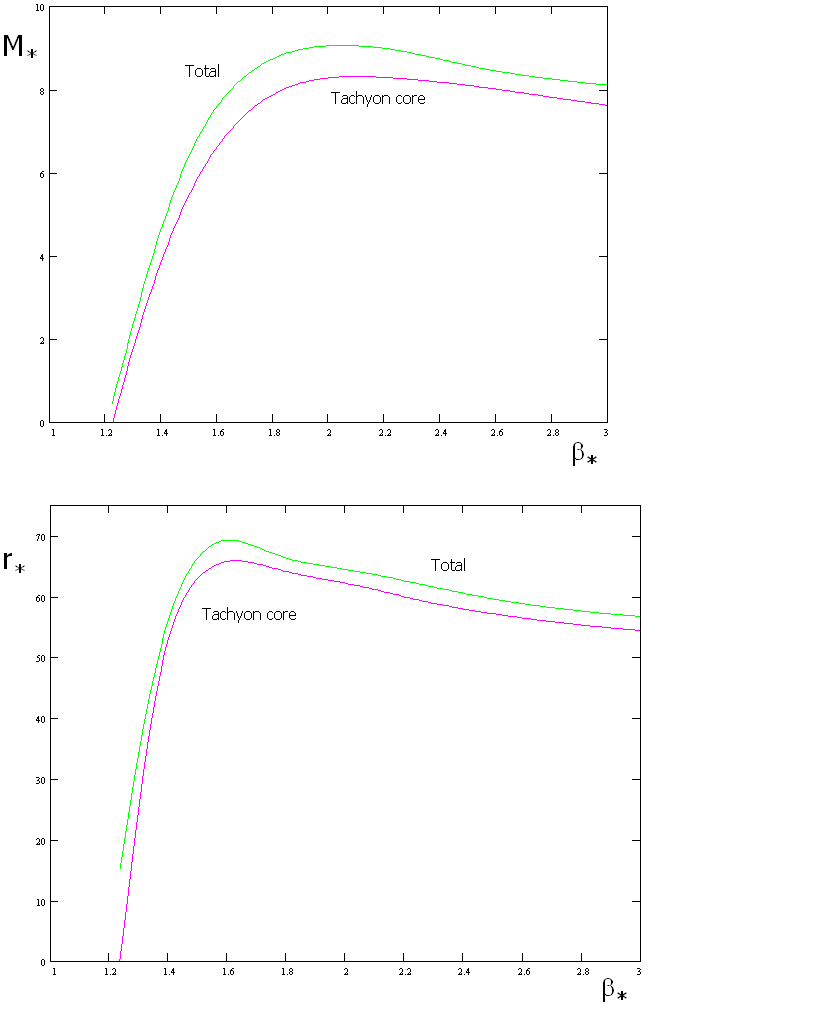}}
\par
\end{figure}

\begin{figure}[tbp]
\caption{The mass and radius of a star with tachyon core in PN envelope vs
variable $\beta_*$ (at $m=138\, \mathrm{MeV}$).}
\label{tsa138a}{\includegraphics[scale=0.7]{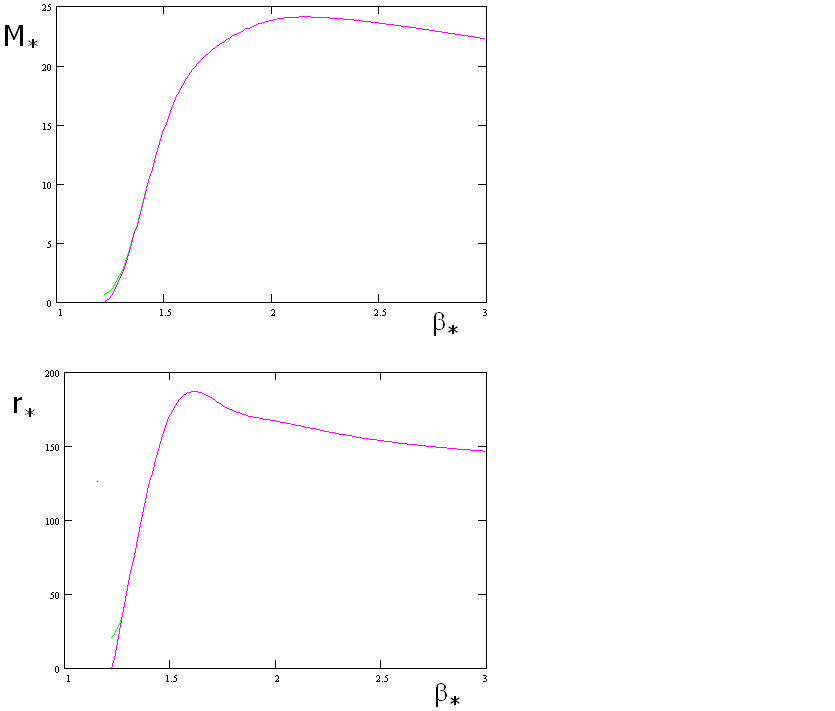}}
\end{figure}

When the tachyon core exists at $\beta _{*}=k_F/m>\sqrt{3/2}$, the total
stellar mass $M_{*}$ and radius $r_{*}$ are greater than those of
non-tachyon star. At large $m$ the main contribution to the total stellar
mass is due to the non-tachyonic envelope, and the sort of the envelope is
very important here. The tachyon core is relatively small when $m>500\,%
\mathrm{MeV}$ because the critical pressure $P_T$ (\ref{p2t}) and critical
density $\rho _T$ (\ref{m2t}) are very large and the tachyon matter is
absent in the peripheral region, being concentrated in the very center (Fig.~%
\ref{tsa939a}-\ref{tsa666a}). As the tachyon mass is chosen small, the
critical pressure $P_T$ (\ref{mt}) is also small, and the tachyon core is
larger and more massive (Fig.~\ref{tsa400a}-\ref{tsa233a}). At very small $%
m<200\,\mathrm{MeV}$ the critical pressure $P_T$ is negligible and the
tachyon core occupies almost the whole volume of the star, while the
envelope is light and thin (Fig.~\ref{tsa138a}).

\begin{figure}[tbp]
\caption{The mass and radius of tachyon core $\bar M_T\left[ \beta
_{*}\right] $ at $m=939\, \mathrm{MeV}$.}
\label{tsa138b}{\includegraphics[scale=0.8]{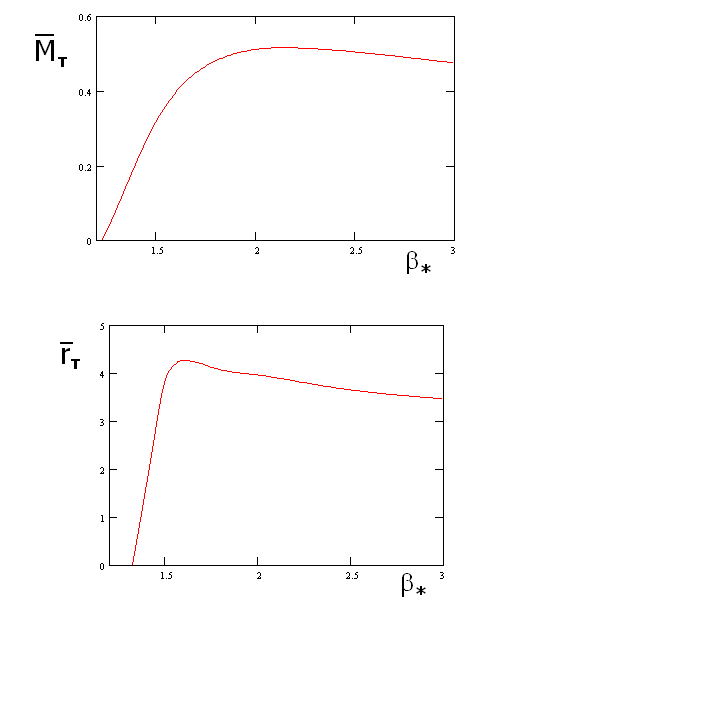}}
\par
\end{figure}

\begin{figure}[tbp]
\caption{The redshift at the surface of the tachyon core vs $\beta_*$.}
\label{tsa88}{\includegraphics[scale=0.7]{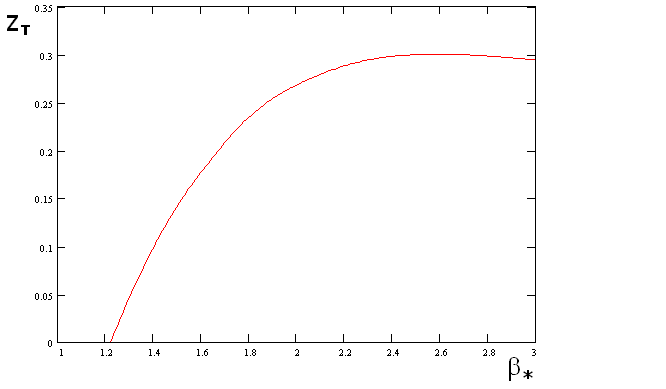}}
\end{figure}

Although the parameters of the whole star are much different, the parameters
of tachyon core do not depend on the properties of the envelope and all
profiles of mass-density $M_T\left[ \beta _{*}\right] $ and radius-density $%
r_T\left[ \beta _{*}\right] $ obey the scaling (\ref{sc33}), see (Fig.~\ref
{tsa138b}).The redshift at the surface of tachyon core is given by universal
dependence (see, Fig.~\ref{tsa88}) which is independent of $m$ and attains
maximum value $z_{\max }\simeq 0.3$ at $\beta _{*}\simeq 2.59$. So, it is
enough to perform calculation at a given $m$ with an arbitrary envelope, and
all parameters of the tachyon core with arbitrary $m$ will be immediately
determined by scaling (\ref{sc33}). Particularly, the maximum mass and
maximum radius are calculated by formula 
\begin{equation}
M_{T\max }=0.52M_{\odot }\frac{m_p^2}{m^2}\qquad r_{T\max }=4.07\,\mathrm{km}%
\frac{m_p^2}{m^2}  \label{sc99}
\end{equation}

As soon as the tachyon core appears at $\beta _{*}=\sqrt{3/2}$, its mass $%
M_T $ and radius $r_T$ rapidly increase with increasing $\beta _{*}$ until
the maximum values are reached, after which the mass and radius decrease at
a slow rate. For each $m$ the mass of tachyon core achieves its maximum
value $M_{T\max }$\textrm{\ }at certain $\beta _{*}\simeq 2.18$ and the
radius of tachyon core achieves its maximum value $r_{T\max }$\textrm{\ }at
certain $\beta _{*}\simeq 1.61$ (which are the same at any $m$). The density 
$\rho _{*\max }$ corresponding to $\beta _{*}\simeq 2.18$, according to (\ref
{m2}) and (\ref{m2t}), is estimated so 
\begin{equation}
\frac{\rho _{*\max }}{\rho _0}\simeq 113\bar m^4\simeq 5.8\rho _T  \label{ma}
\end{equation}

According to formula (\ref{sc99}), we find that 
\begin{equation}
M_{T\max }\simeq 2000M_{\odot }\qquad r_{T\max }\simeq 18,000~\mathrm{km}
\label{sc6}
\end{equation}
when the tachyon mass is $m=14\,\mathrm{MeV}$ and the tachyon critical
density is $\rho _T=5\times 10^{-7}\rho _0=1.6\times 10^8\,\mathrm{g\cdot cm}%
^{-3}$, while 
\begin{equation}
M_{T\max }\simeq 4\times 10^{22}M_{\odot }\qquad r_{T\max }\simeq 4\times
10^{22}~\mathrm{km}  \label{sc6b}
\end{equation}
when the tachyon mass is $m=10^{-11}m_p=0.01\,\mathrm{eV}\sim 100^{\circ }K$
and the tachyon critical density is $\rho _T=3\times 10^{-30}\,\mathrm{%
g\cdot cm}^{-3}$. The latter value is not so far from the well known
observation data about the Universe, and the results of the present paper
may find further application to cosmology. The most important property of
tachyonic self-gravitating body is that it can exist being embedded in some non-tachyonic medium (whose parameters do not influence the mass and radius of the tachyon core) and that its maximum mass and radius are estimated according to formula (\ref{sc99}).

The author is grateful to Erwin Schmidt for discussions.

\begin{thebibliography}{99}
\bibitem{S2002}  A. Sen, JHEP \textbf{0207}, 065 (2002). \texttt{%
arXiv:hep-th/0203265}

\bibitem{BBS2003}  M. C. Bento, O. Bertolami, and A. A. Sen, Phys. Rev. D 
\textbf{67}, 063511 (2003). \texttt{arXiv:hep-th/0208124}

\bibitem{FKS02}  A. Frolov, L. Kofman, and A. Starobinsky, Phys.Lett. B 
\textbf{545}, 8 (2002). \texttt{arXiv:hep-th/0204187}

\bibitem{D1}  E. J. Copeland, M. R.Garousi, M. Sami, S. Tsujikawa, Phys.
Rev. D \textbf{71}, 043003 (2005). \texttt{arXiv:hep-th/0411192}

\bibitem{D2}  U. Debnath, Class. Quant. Grav. \textbf{25}, 205019 (2008). 
\texttt{arXiv:0808.2379v1 [gr-qc]}

\bibitem{M84}  St. Mr\'owczy\'nski, Nuovo Cim. B \textbf{81}, 179 (1984).

\bibitem{DHR89}  R. L. Dawe, K. C. Hines and S. J. Robinson, Nuovo Cim. A 
\textbf{101}, 163 (1989).

\bibitem{KRS07}  K. Kowalski, J. Rembielinski, and K.A. Smolinski, Phys.
Rev. D \textbf{76}, 045018 (2007). \texttt{arXiv:0712.2725v2 [hep-th]}

\bibitem{TV2011c}  E. Trojan and G. V. Vlasov, Phys. Rev. D \textbf{83},
124013 (2011). \texttt{arXiv:1103.2276 [hep-ph]}

\bibitem{TV2010}  E. Trojan and G.V. Vlasov, Phys.\ Rev. C \textbf{81},
048801 (2010).

\bibitem{V98}  G.V. Vlasov, Phys.\ Rev. C \textbf{58}, 2581 (1998).

\bibitem{W90}  J. D. Walecka, \textit{Theoretical nuclear and subnuclear
physics}, 2nd. ed. (World Scientific, 2004), p. 19.

\bibitem{T2011j}  E. Trojan, \textit{Careful calculation of thermodynamical
functions of tachyon gas}. \texttt{arXiv:1109.10261 [astro-ph.CO]}

\bibitem{OV39}  J. R. Oppenheimer and G. M. Volkoff, Phys. Rev. \textbf{55},
374 (1939).

\bibitem{RR74}  C. E. Rhoades and R. Ruffinin, Phys. Rev. Lett. 32, 324
(1974).

\bibitem{KB96}  V. Kalogera and G. Baym, ApJ \textbf{470} L61 (1996). 
\texttt{arXiv:astro-ph/9608059}

\bibitem{BPS75}  G.Baym, C. Pethick, and P. Sutherland, ApJ, \textbf{170},
299 (1971).

\newpage
\end{thebibliography}
\end{document}